# Struggle for Existence:

# the models for Darwinian and non-Darwinian selection

## G. Karev[1], F. Berezovskaya[2]

[1]National Center for Biotechnology Information, Bethesda, MD 20894, USA.
Email: karev@ncbi.nlm.nih.gov
[2] Howard University, Washington, DC 20059, USA
e-mail: fberezovskaya@howard.edu

**Abstract**

Classical understanding of the outcome of the struggle for existence results in the Darwinian "survival of the fittest". Here we show that the situation may be different, more complex and arguably more interesting. Specifically, we show that different versions of inhomogeneous logistic-like models with a distributed Malthusian parameter imply non-Darwinian "survival of everybody". In contrast, the inhomogeneous logistic equation with distributed carrying capacity shows Darwinian "survival of the fittest". We also consider an inhomogeneous birth-and-death equation and give a simple proof that this equation results in the "survival of the fittest". In addition to this known result, we find an exact limit distribution of the parameters of this equation. We also consider "frequency-dependent" inhomogeneous models and show that although some of these models show Darwinian "survival of the fittest", there is not enough time for selection of the fittest species. We discuss the well-known Gauze's Competitive exclusion principle that states that "Complete competitors cannot coexist". While this principle is often considered as a direct consequence of the Darwinian "survival of the fittest", we show that from the point of view of developed mathematical theory complete competitors can in fact coexist indefinitely.

1. **Introduction. Darwinian evolution and Competitive Exclusion principle**

In his 1934 book, G.F. Gause described his experiments where he cultivated two types of yeast, *S. cerevisiae* and *S. kephir*, to experimentally test the mathematical theory of struggle for existence developed mainly by Lotka and Volterra. He developed this experiment to test mathematical models of competition between organisms with a limited food supply.

Gause considered his experiments as experimental proof of an important theoretical result that followed from mathematical models, namely that two species with similar ecology cannot live in the same region. This statement is now well-known in ecology as the "Competitive exclusion principle" or as Gause or Volterra-Gause principle.

Hardin (1960) reformulated the "Exclusion principle" in a more aphoristic form: "Complete competitors cannot coexist" (and noticed that every one of the four words is ambiguous).

Later this statement was generalized to the case of community consisting of an arbitrary number of species: "no stable equilibrium is possible if some $r$ species are limited by less than $r$ resources" (Levin 1970). It was assumed here that the growth rates of species depend linearly on the resources. Recently it was shown (Szilagyi, Zachar and Szathmary, 2013) that the principle of competitive exclusion holds for template replicators if resources (nucleotides) affect growth linearly. It is important to notice that the assumption that the growth rates of species are linear functions of resources is crucial; if this assumption is relaxed then coexistence of $r$ species on $k < r$ resources is possible (perhaps, not with constant densities), see an important paper by Armstrong, McGehee (1980) and references therein.

In an interesting essay Hardin (1960), citing Gilbert et al. (1952) stated that Gause "draws no general conclusions from his experiments, and moreover, makes no statement which resembles any wording of the hypothesis which has arisen bearing his name". Furthermore, "How curious it is that the principle should be named after a man who did not state it clearly, who misapprehended its relation to theory, and who acknowledged the priority of others!"

These statements seem to be incorrect. Indeed, the main focus in the Gause's book was concentrated on an experimental study of the struggle for existence. At the same time, he derived equations for the struggle of existence that "express quantitatively the process of competition between two species for the possession of a certain common place in the microcosm" of the following form (see Eq. (12) in Gause (1934)):

$$\frac{dN_1}{dt} = b_1 N_1 (1 - \frac{N_1 + \alpha N_2}{K_1}), \tag{G}$$

$$\frac{dN_2}{dt} = b_2 N_2 (1 - \frac{N_2 + \beta N_1}{K_2}).$$

He noticed that this equation formally coincides with known Volterra' equation "but it does not include any parameters dealing with the food consumption, and simply expresses the competition between species in terms of the growing populations themselves."

He emphasized in Ch. 3 (Gause, 1934), that equation (G) "does not permit of any equilibrium between the competing species occupying the same "niche," and leads to the entire displacement of one of them by the other. This has been pointed out by Volterra, Lotka and even earlier by Haldane, and for the experimental confirmation and a further analysis of this problem the reader is referred to Chapter V. …The mathematical considerations show that with *usual* $\alpha$ and $\beta$, there cannot simultaneously exist positive values for population sizes of both (species). One of the species must eventually disappear. This apparently harmonizes with the biological observations. As we have pointed out in Chapter II, both species survive indefinitely only when they occupy different niches in the microcosm in which they have an advantage over their competitors".

It looks like a clear formulation of the exclusion principle.

Although this principle seems to follow directly from Darwinian natural selection, there are some problems with it that have been later identified, such as the "paradox of plankton". The diversity of natural phytoplankton seems to contradict the competitive exclusion principle. Although most algae compete for the same inorganic nutrients, often more than 30 species coexist even in small parcels of water.

Hutchinson (1961) posed his classic question: "How is it possible for a number of species to coexist in a relatively isotropic or unstructured environment, all competing for the same sorts of materials?" This problem was discussed in many papers; in particular, Wilson (1990) wrote: "The almost ubiquitous existence of multi-species communities is one of the few firm facts in ecology. How can alpha species diversity be as high as it is within most actual communities, in the face of the Principle of Gause that no two species can permanently occupy the same niche? Why does competitive exclusion not occur, leaving only one species - the one with the highest competitive ability?"

Wilson also noticed that a similar question exists for tropical rain forest and coral reef communities. He proposed 12 possible mechanisms to explain the paradox:

1. Niche Diversification; 2. Pest Pressure; 3. Equal Chance; 4. Gradual Climate Change;
5. Intermediate-timescale Disturbance; 6. Life History Differences;
7. Initial Patch Composition; 8. Spatial Mass Effect; 9. Circular Competitive Networks;
10. Cyclic Succession; 11. Aggregation; 12. Stabilizing Coevolution.

Several other hypotheses have since then been proposed to address the problem of the Exclusion principle, such as:

1) Resource models of coexistence (Levin 1970; Armstrong and McGehee 1980);
2) Biodiversity under non-equilibrium conditions by species oscillations and chaos (Huisman & Weissing 1999);
3) Neutral theory: evolution of ecological equivalence or niche convergence (Hubbell 2001, 2006).

Additionally, some of the modern theories suggest that the coexistence of species with similar competitive abilities can co-occur in nature as a result of a balancing act between fitness equalizing processes such as tradeoffs and fitness stabilizing processes like the rare species advantage.

One can reasonably ask: if there are so many exclusions from the exclusion principle, then maybe something is wrong either with the ways the principle was tested, or maybe with the principle itself?

An interesting and important point was suggested by Hardin (1960): "There are many who have supposed that the principle is one that can be proved or disproved by empirical facts, among them Gause himself. Nothing can be farther from the truth... The "truth" of the principle is and can be established only by theory, not being subject to proof or disproof by facts…Indeed, let two non-interbreeding species that seem to have the same ecological characteristics be placed in the same location; if one of species extinguished the other, one says that the principle is proved. But if the species continue to coexist indefinitely, one may decide that there must be some subtle difference in ecology."

We believe that the "theory" here means the mathematical models of selection and struggle for existence, in accordance with Gause himself. The Gause'principle can be considered as a particular case (or consequence) of the Darwinian "selection of the fittest". Conversely, a common opinion is that the struggle for existence results in the Darwinian "survival of the fittest". In what follows we show that the situation is different and more complex.

We will consider a series of mathematical models to study what kind of selection follows from the models.

But first, let us discuss the Gause's equation (G). As Gause noticed, with *usual* α and β, one of the species must eventually disappear and the model shows Darwinian selection of the fittest. Gause did not explain what the *usual* values of α and β are. Now one can answer this question completely as the model (G) has been studied at full, see, e.g., (Bazykin 1998).

Let us denote $K = K_1/K_2$; let $(N_1, N_2)$ be the numbers of the species. Then:

1) if $\alpha > K, \beta < 1/K$, then (0,1) is the only stable equilibrium; 2nd species dominates;
2) if $\alpha < K, \beta > 1/K$, then (1,0) is the only stable equilibrium; 1st species dominates;
3) if $\alpha > K, \beta > 1/K$, then any species can dominate dependently on initial conditions;
4) if $\alpha < K, \beta < 1/K$, then there exists a positive stable equilibrium
$$N_1 = \frac{K_1 - \alpha K_2}{1 - \alpha\beta}, \quad N_2 = \frac{K_2 - \beta K_1}{1 - \alpha\beta}.$$

The last condition means that the intensity of intra-species competition is less than the intensity of inter-species interactions, and in this case, **the species coexist forever**. So, depending on the parameters that characterize the niche, either one or both species may survive. The result does not depend on Malthusian growth rates of each species. Hence, the Gause model (G) *does not* imply the exclusion principle for a large domain of the model parameters (see Fig.1 for a parametric portrait of the model (G)).

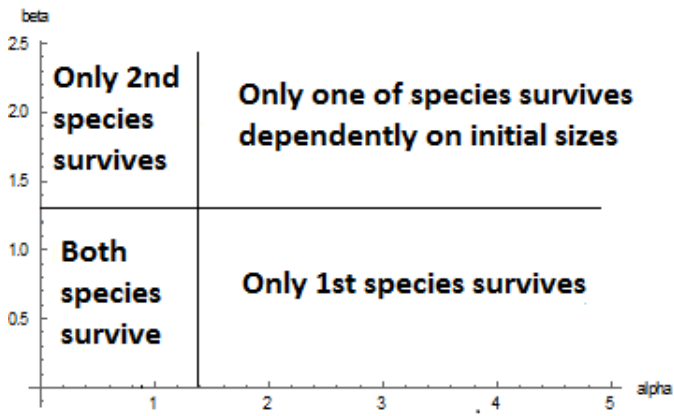

**Figure 1.** Schematic parametric portrait of Gause's model, as given by System (G).

Let us consider a numerical example that shows the coexistence of both species (domain 4)): $K_1 = K_2 = 1$, $\alpha = 0.2, \beta = 0.1$ in System (G). In this case, the limit equilibrium values are $N_1 = 0.8163$, $N_2 = 0.9184$. The dynamics of number of the species is shown in Fig.2.

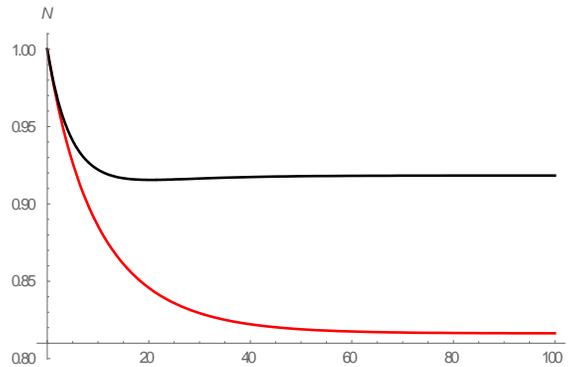

**Figure 2**. Graphs of $N_1(t)$ (red) and $N_2(t)$ (black) as defined in System (G).

Remark, that existence of the 3rd domain shows perhaps the most striking deviation from the Competitive principle as two species are complete competitors, but each can win. In experiments described in (Park, Lloid, 1995) indeed one of the species was completely eliminated but it was not always the same one.

Hardin (1960) suggested the following refinement of the Exclusion principle:

i. if two non-interbreeding populations occupy precisely the same ecological niche,
ii. if they occupy the same geographical territory,
iii. if population A multiplies even a bit faster than population B,

then ultimately A will completely displace B, which will become extinct.

However, the exclusion principle even in this form also does not follow from mathematical modeling. Let us consider again the Gause model (G) with $\alpha = \beta = 1$. This simple model (that coincides with the standard Volterra model of competition) satisfies the Hardin conditions if $b_1 \neq b_2$. It is easy to show that neither species becomes extinct; a numerical example is shown in Fig.3.

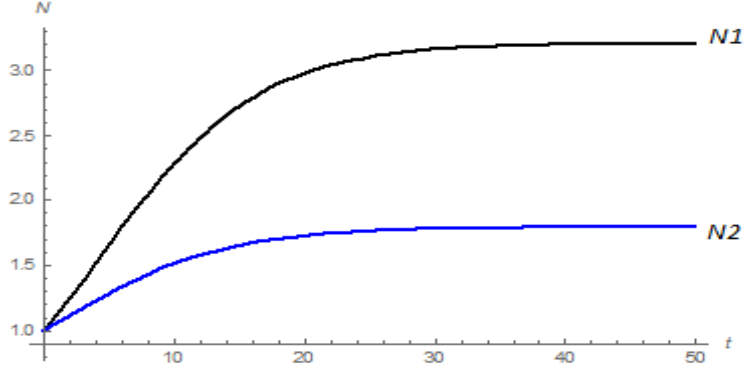

**Figure 3**. Solution to equation (G) with $\alpha = \beta = 1, K_1 = K_2 = 1, b_1 = 0.2, b_2 = 0.1$

Before we proceed further and consider models appropriate for studying the exclusion principle and, more generally, the outcomes of natural selection, let us emphasise that natural selection can operate only if the community is composed from many species (at least, initially) or if the evolving population is non-homogeneous. Mathematical frameworks for studying the dynamics of inhomogeneous populations and communities were developed in (Karev 2010a,b).

The simplest conceptual model describing the Darwinian "survival of the fittest" is the inhomogeneous Malthusian model of a population, composed from clones $l(t,a)$, where $a$ is the Malthusian growth rate per individual in the clone,

$$\frac{dl(t,a)}{dt} = al(t,a). \tag{1.1}$$

The solution to this equation is $l(t,a) = l(0,a)\exp(at)$; if $a_1 < a_2$, then $\frac{l(t,a_1)}{l(t,a_2)} = \frac{l(0,a_1)}{l(0,a_2)} e^{(a_1-a_2)t} \to 0$ as $t \to \infty$. Hence, every clone with smaller growth rate $a$ will be overtaken over time by a clone with a larger growth rate. It means the "survival of the fittest" in the simplest (and actually trivial) form. Notice that the selection of the fittest in population (1.1) can be realized only if the size of each clone increases indefinitely. Hence, inhomogeneous Malthusian model (1.1) shows the "survival of the fittest" only when it becomes unrealistic and cannot be considered as a sort of "mathematical justification" of the Darwinian evolution.

In what follows we study different models of inhomogeneous populations of "Malthusian" type. These models describe a population composed of individuals with different reproduction rates (Malthusian parameters) $a$; we refer to the set of all individuals with the given value of parameter $a$ as an $a$-clone. Let $l(t,a)$ be the size of $a$-clone at the moment $t$. We assume that the growth rate of each clone depends on the total population size $N(t)$. Dynamics of such a population can be described by the following model:

$$\frac{dl(t,a)}{dt} = al(t,a)g(N), \quad N(t) = \int_A l(t,a)da \tag{1.2}$$

where $g(N)$ is some function, chosen depending on the specifics of each model. For example, if $g(N) = const$, then (1.2) is an inhomogeneous Malthusian model (1.1); if $g(N) = (1 - \frac{N}{C})$, then (1.2) describes an inhomogeneous logistic model, where each clone grows logistically with a common carrying capacity $C$. Different versions of inhomogeneous logistic equations are considered below in details. We refer to models (1.2) as inhomogeneous *density-dependent* models (D-models for brevity) as the right hand of the equation is proportional to the clone density $l(t,a)$. We will consider also *frequency-dependent* models (F-models for brevity) that have the form

$$\frac{dl(t,a)}{dt} = aP(t,a)g(N) \tag{1.3}$$

where $P(t,a) = l(t,a)/N(t)$ is the frequency of the parameter $a$.

Notice that formally model (1.3) is a special case of (1.2), but in applications it shows some interesting additional properties, see (Karev, 2014). It was shown in (Kareva, Karev, 2017) that many experimental growth curves, including non-standard two- and three-stage growth curves can be understood and described within the frameworks of F-models.

The problem of different possible outcomes of selection was discussed in (E. Szathmary and M. Smith, 1997; see also Szathmary & Gladkih, 1989, and Szathmary. 1991); they represent the model of pre-biological evolution of replicators by the equation for the concentration of molecules

$$\frac{dx_i}{dt} = k_i x_i^p. \tag{1.4}$$

Here $x_i$ is the amount or concentration of replicator type $i$, and $k_i$ is the analogue to (Malthusian) growth rate. In addition to the standard exponential model with $p = 1$, they also considered the super-exponential or "hyperbolic" equation with $p > 1$ and the sub-exponential or "parabolic" equation with $p < 1$, see examples and references in (Szathmary & Smith, 1997).
In the *hyperbolic* case, the "most common" replicator (i.e. the replicator with the largest value $k_i x_i(0)$) becomes dominant, implying "*survival of the common*".

In the *parabolic* cases, the ratio between the concentrations remains finite, implying "*survival of everybody*".

The last property was, perhaps, the main reason why the models of populations composed from "parabolic" clones attracted attention of many authors. The model is unrealistic at small community density because the growth rate per individual tends to infinity as the density tends to 0. It is the unique mathematical reason why the model shows the "survival of everybody". As for the hyperbolic case, the growth rate per individual and the population size increase indefinitely at a finite time moment. So, the deviations from the Darwinian selection in both cases start when the models become unrealistic.

A possible more realistic explanation of non-linear population growth (1.4) within the frameworks of inhomogeneous *frequency-dependent* models was suggested in (Karev, 2014). Interestingly, it appears to have been underappreciated that a much more realistic model, the inhomogeneous *density-dependent* logistic equation with distributed Malthusian parameter, also shows the "survival of everybody". This equation presents a simple conceptual model for Malthusian Struggle for Existence, which accounts for both free exponential growth and for resource limitations:

$$\frac{dl(t;a)}{dt} = al(t;a)(1 - \frac{N}{C}) , \qquad (1.5)$$

$N(t) = \int_A l(t,a) da.$

Here $a \in A$ is the Malthusian reproduction rate, which is assumed to be distributed with initial distribution $P(0,a)$, $C = const$ is the common caring capacity, and $N$ is the total population size. Let us cite Gause (1934): "All populations have the capacity to grow exponentially under ideal conditions, and no population can grow exponentially forever – there are limits to growth. This generates the Malthusian Struggle for Existence". One may recognize here the description of logistic population model (1.5), but this model *does not* generate the Malthusian struggle for existence and survival of the fittest; see the next section for details.

On the other hand, Achleh et.al. (1999) considered an inhomogeneous logistic equation in the form of birth-and-death equation

$$\frac{dl(t;b.d)}{dt} = l(t;b,d)(b - dN) \qquad (1.6)$$

where $b, d$ are the birth and death per capita rates correspondingly of the clone $l(t; b, d)$. Evidently, equations (1.5) and (1.6) are equivalent if the parameters $a, C$ in (1.5) and $b, d$ in (1.6) are constants, but it is not so if the parameters are distributed. It was proven in (Achleh et al, 1999) and the following papers of Ackleh and coauthors that under general conditions only "the fittest", i.e. those individuals that have the largest value of $b$ and the smallest value of $d$, survive in the population (1.6). Notice, that the "survival of everybody" for inhomogeneous logistic equation (1.5) with distributed Malthusian parameter and fixed carrying capacity $C$ easily follows from the results of Ackleh et al. (1999).

In what follows we study different generalizations of "density-dependent" inhomogeneous logistic equation with distributed Malthusian parameter and show that all of them result in "survival of everybody". In contrast, the inhomogeneous logistic equation with distributed carrying capacity shows Darwinian "survival of the fittest". We consider also inhomogeneous birth-and-death equation in the form (1.6) and give a simple proof that this equation results in "survival of the fittest". In addition to this known result, we found an exact limit distribution of the parameters of this equation. We also consider "frequency-dependent" inhomogeneous models and show that although some of these models show Darwinian survival of the fittest, there is not enough time for selection of the fittest species.

## 2. *Solution to the inhomogeneous logistic equation*

Inhomogeneous logistic equation (1.5) with distributed Malthusian parameter $a$ can be solved with the help of HKV method (after <u>h</u>idden <u>k</u>eystone <u>v</u>ariables) (Karev, Kareva 2014), which follows from the Reduction theorem (Karev, 2010). A simplified version of the method is described in Appendix 1.

We suppose that in equation (1.5) the initial population size $N(0)$ is less than the carrying capacity $C$, i.e. $N(0) < C$. All equations (2.1)-(2.6) below are particular cases of corresponding equations given in Mathematical Appendix 1.

Define the auxiliary keystone variable $q(t)$ by the equation

$$\frac{dq}{dt} = (1 - \frac{N}{C}), q(0) = 0. \qquad (2.1)$$

Then

$$l(t, a) = l(0, a) \exp(aq(t)) \qquad (2.2)$$

Define the mgf (moment generating function) $M(\lambda)$ of the initial distribution of the parameter $a$,

$M(\lambda) = \int_A \exp(\lambda a) P(0,a) da.$

Then the total size of the population is given by the formula

$$N(t) = N(0)M(q(t)). \tag{2.3}$$

The equation for the auxiliary variable can be now written in a closed form and solved:

$$\frac{dq}{dt} = 1 - \frac{N(0)}{C}M(q), \quad q(0) = 0. \tag{2.4}$$

The population size solves the logistic-like equation

$\frac{dN}{dt} = E^t[a]N(1 - \frac{N}{C})$

where $E^t[a] = \int_A a P(t,a) da$ is the current mean value of the parameter $a$.

The current distribution of the parameter $a$ is given by the formula

$$P(t,a) = \frac{l(t,a)}{N(t)} = \frac{\exp(aq(t))}{M(q(t))} P(0,a). \tag{2.5}$$

The current mean value of the parameter $a$ can be easily computed by the formula

$$E^t[a] = \frac{d\ln M}{dq}(q(t)). \tag{2.6}$$

Hence, we have reduced inhomogeneous many- or infinitely-dimensional logistic equation (1.5) to a single equation (2.4) for $q(t)$.

The keystone variable $q(t)$ can be considered as the "internal time" of the population: the dynamics of inhomogeneous logistic model (1.5) with respect to the internal time is identical to the dynamics of the inhomogeneous Malthusian model with respect to the regular time (see Appendix 1). The principle difference is that for inhomogeneous logistic model $q(t)$ tends to a finite value $q^* < \infty$ as $t \to \infty$, which is the single equilibrium of equation (2.4) and can be found as the solution to the equation $\quad M(q^*) = C/N(0)$.

**Proposition 1**. The single equilibrium $q^*$ of equation (2.4) is stable for any initial distribution of the Malthusian parameter $a$.

Indeed, $\frac{dq}{dt} = 0$ if and only if $1 - \frac{N}{C} = 1 - \frac{N(0)}{C}M(q) = 0$. The function $M(q) \geq 1$ and increases monotonically; hence the equation $1 - \frac{N(0)}{C}M(q) = 0$ has a single solution $q^*$. The function $q(t)$ also increases monotonically, because $q(0) = 0, M(0) = 1, \frac{N(0)}{C} < 1$, hence

$\frac{dq}{dt} = 1 - \frac{N(0)}{C}M(q(t)) > 0$ until $N(0)M(q(t)) = N(t) < C$.

Next, $\frac{d}{dq}\left(1 - \frac{N(0)}{C}M(q)\right) = -\frac{N(0)}{C}\frac{dM}{dq} < 0$, hence the equilibrium $q$ is stable, as desired.

**Example.** Let the initial distribution of the Malthusian parameter be exponential, $P(0, a) = \exp(-ma)/m$, $m = const > 0$; its mgf $M(\lambda) = m/(m - \lambda)$. Then

$$\frac{dq}{dt} = 1 - \frac{N(0)}{C}\frac{m}{m-q}. \tag{2.7}$$

The following figure shows the solution to this equation as $m = 1, \frac{N(0)}{C} = 0.1$.

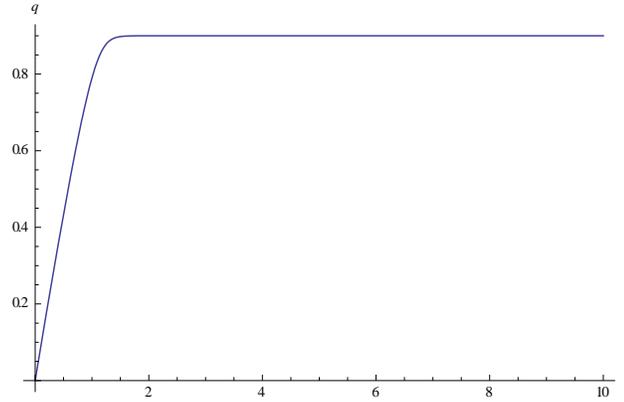

**Figure 4**. The "internal time" $q(t)$ against the plain time $t$ for equation (2.7); $q(t) \to q^* = 0.9$ as $t \to \infty$.

Now we can see, that the limit state of inhomogeneous logistic model (1.5) coincides with the current state of the inhomogeneous Malthus model at the instant $q^*$. The limit stable population size and the limit distribution of the parameter $a$ are

$$N^* = N(0)M(q^*), \tag{2.8}$$
$$P^*(a) = P(0; a)\exp(q^*a)M(q^*).$$

Let us emphasize a notable property of the inhomogeneous logistic model (1.5) with a distributed Malthusian parameter: it remains inhomogeneous at any instant and has a non-trivial limit distribution of the parameter at $t \to \infty$. Every clone that was present initially will be present in the limit stable state. Therefore, inhomogeneous logistic model illustrates the phenomenon of "survival of everybody" in the population, in contrast to Darwinian "survival of the fittest". The Exclusive principle does not hold for populations that can be described by this model.

## 3. Generalized logistic inhomogeneous models with distributed Malthusian parameter

Different generalization of the standard logistic equation are known in literature, which were applied to many particular problems, see (Tsoularis, Wallace, 2002). Most of them are particular cases of the generalized logistic equation

$$\frac{dN}{dt} = rN^\alpha [1 - \left(\frac{N}{C}\right)^\beta]^\gamma \qquad (3.1)$$

where $\alpha, \beta, \gamma$ are the model parameters. Examples are the Blumberg equation (Blumberg, 1968) with $\beta = 1$, the Richards equation (Richards, 1959) with $\alpha = \gamma = 1$, von Bertalanffy equation (Von Bertalanffy, 1938) with $\alpha = \frac{2}{3}, \beta = \frac{1}{3}, \gamma = 1$.

The simplest (but not unique) inhomogeneous version of model (3.1) is

$$\frac{dl(t,a)}{dt} = al(t,a)N^{\alpha-1}[1 - \left(\frac{N}{C}\right)^\beta]^\gamma \qquad (3.2)$$

so that

$$\frac{dN}{dt} = E^t[a]N^\alpha [1 - \left(\frac{N}{C}\right)^\beta]^\gamma. \qquad (3.3)$$

Again, $l(t, a)$ is the density of the clone, having intrinsic Malthusian parameter $a$; the expression $F(N) = N^{\alpha-1}[1 - \left(\frac{N}{C}\right)^\beta]^\gamma$ describes the dependence of the growth rate of each clone on total population size.

The method of solution to the multi-dimension equation (3.2) is similar to that used in s.2 for solution to the inhomogeneous logistic equation (1.5).

Define the keystone variable (the internal time of the population) $q(t)$ by the equation

$$\frac{dq}{dt} = F(N) \equiv N^{\alpha-1}[1 - \left(\frac{N}{C}\right)^\beta]^\gamma, q(0) = 0. \qquad (3.4)$$

Then

$$l(t, a) = l(0, a)\exp(aq(t)). \qquad (3.5)$$

The total size of the population $N(t)$ and the current distribution $P(t, a)$ of the parameter $a$ are given by the formulas (see equations (A1.7) and (A1.9))

$$N(t) = \int_A l(t, a)da = N(0)M(q(t)),$$

$$P(t, a) = \frac{l(t,a)}{N(t)} = \frac{\exp(aq(t))}{M(q(t))} P(0, a)$$

where $M(\lambda)$ is the mgf of the initial distribution $P(0, a)$.

The equation for $q(t)$ can be written in a closed form:

$$\frac{dq}{dt} = M(q)^{\alpha-1}[1 - \left(\frac{M(q)}{C}\right)^\beta]^\gamma, \quad q(0) = 0. \tag{3.6}$$

By this way, the inhomogeneous multi- or infinitely-dimensional logistic equation (3.2) is reduced to a single equation (3.6) for $q(t)$. Now all statistical characteristics of the model (such as the current mean value and variance of the parameter $a$) can be effectively computed.

Figure 5 shows the behavior of $q(t)$ for different versions of the generalized logistic equation; the initial distribution of the Malthusian parameter $a$ was exponential with the mgf $M(q) = \frac{1}{1-q}$.

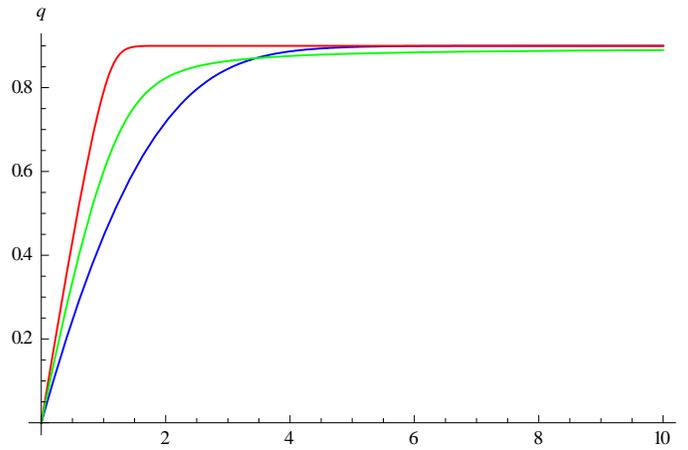

**Figure 5.** The plots of internal time $q(t)$ for von Bertalanffy equation (blue), Richards equation with $\gamma = 3$ (green), Richards equation with $\gamma = 1$ (red); in all equations $K = 10$, $M(q) = \frac{1}{1-q}$.

We can see, that behaviors of $q(t)$ are very similar for all three equations and hence the behaviors of its solutions are also very close due to formula (3.5). The limit value $q^*$ solves the equation $\frac{M(q)}{C} = 1$ and is the same for *all generalized logistic equations* (3.2) with given caring capacity $C$ and given mgf of the initial distribution; $q^*$ is the single equilibrium of the equation (3.6). It is easy to show that this equilibrium is stable as the derivative of the right hand of equation (3.6) is negative as $q = q^*$. The limit distribution of inhomogeneous model (3.1) coincides with the current distribution of inhomogeneous Malthus model (1.1) at the instant $q^*$. Hence, all clones that were present in the population at the initial time moment will have positive frequency in the limit state of the population. It means "survival of everybody" for all inhomogeneous generalized logistic models (3.1). Hence, the Exclusion principle is not valid for communities, composed from species that grow according to generalized logistic equation (3.2).

In this and previous sections we have considered inhomogeneous versions of different logistic-like models. It was assumed that the population is composed from logistic-like clones (species) that grow according to equations (1.5) or (3.2). In all cases, the equations for total population size differ from the original logistic equations, e.g., compare equations (3.1) and (3.3); so the dynamics of its solutions is also different. Let us consider, for example, the standard logistic equation and its inhomogeneous modification, i.e. equations (3.1) and (3.3) as $\alpha = \beta = \gamma = 1$. Let the initial distribution of the parameter $a$ be exponential. Figure 5 allows us to compare the dynamics of total population size for standard and inhomogeneous logistic models with different mean values of the initial distribution. We can see that the dynamics is different, although the limit values of $N(t)$ as $t \to \infty$ are the same.

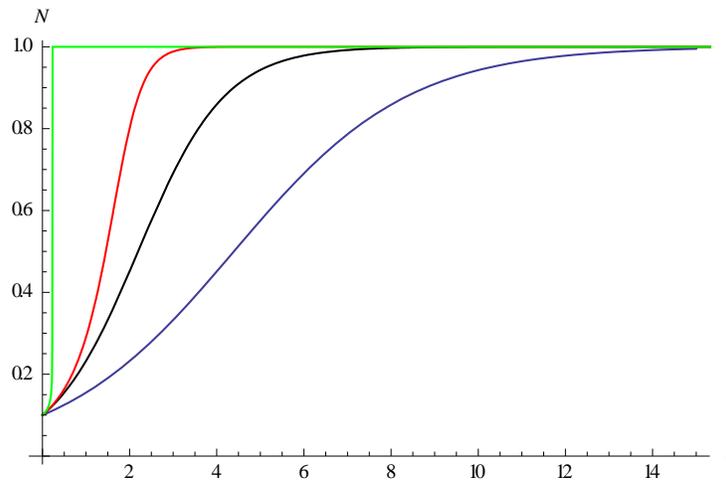

**Figure 6.** Dynamics of the total population size; the standard logistic equation with *a*=0.5 (blue); inhomogeneous logistic equation (1.5) with exponential initial distribution: a) $E^0[a] = 0.001$ (black); b) $E^0[a] = 0.5$ (red); c) $E^0[a] = 5$ (green).

Taking into account that all real populations are non-homogeneous, the question arises: if there exists a model of inhomogeneous population such, that its total population size solves the generalized logistic equation (3.1). The answer is affirmative: it is possible to construct inhomogeneous *frequency-dependent* model, whose total population size solves this equation.

Let us consider inhomogeneous frequency-dependent model

$$\frac{dl(t,a)}{dt} = raP(t,a)[1 - \left(\frac{N}{C}\right)^\beta]^\gamma . \tag{3.7}$$

The following statements are proved in Math Appendix 2:

**Proposition** 2. The total size of inhomogeneous population (3.7) with initial Gamma-distribution (A2.7) of the Malthusian parameter $a$ solves the generalized logistic equation (3.1).

**Corollary 1.** The total size of the inhomogeneous population $\frac{dl(t,a)}{dt} = raP(t,a)(1 - \frac{N}{C})$ with initial exponential distribution of the Malthusian parameter $P(0,a) = \exp\left(-\frac{a}{N(0)}\right)/N(0)$ solves the standard logistic equation $\frac{dN}{dt} = rN(1 - \frac{N}{C})$.

**Corollary 2.** The total size of the inhomogeneous population $\frac{dl(t,a)}{dt} = raP(t,a)$ with initial exponential distribution of the Malthusian parameter $P(0,a) = \exp\left(-\frac{a}{N(0)}\right)/N(0)$ solves the standard Malthusian equation $\frac{dN}{dt} = rN$.

The limit distribution of inhomogeneous model (3.7) also coincides with the current distribution of inhomogeneous Malthus model (1.1) at the instant $q^*$ that solves the equation $\frac{M(q)}{C} = 1$. Hence, all clones (species) that were present in the population at the initial time moment will present forever in the population. The model shows "survival of everybody" and the Exclusion principle is not valid for communities, composed from species that grow according to frequency-dependent model (3.7).

In conclusion of this section let us notice, that it was supposed in logistic-like models studied above that the growth of each clone depended only on total population size $N(t)$. In general, the growth rates of clones may depend also on other population characteristics such as total or average biomass, a territory per individual, etc. These quantities, known also as "regulators" evolve with time providing self-regulation of the system dynamics. As a result, we arrive to "multi-factor" logistic-like models considered in Mathematical Appendix 3. We proved there, that multi-factor models also show "survival of everybody"

## 4. Inhomogeneous Gompertz equation

The Gompertz growth curve (Gompertz, 1825) is widely used in cancer modeling, ecological problems, etc., see, e.g., (Kendal, 1985; Gyllenberg, Webb, 1988; Bajzer, Vuk-Pavlovic, 2000). Statistical analysis of results of Gause experiments (Fig.-s 24, 25 in Gause 1934), provided in (Nedorezov, 2015) showed that in some cases the Gompertz curve describes experimental time series even better than the logistic curve.

The Gompertz curve is given by the equation

$N(t) = N(0)\exp(A(1 - e^{-rt}))$, where $A, r$ are positive parameters.

The Gompertz curve can be also written in equivalent form

$$N(t) = K\exp(\ln\frac{N(0)}{K} e^{-rt}) \qquad (4.1)$$

where $K, r$ are positive parameters.

The curve (4.1) is a solution to the equation

$\frac{dN}{dt} = rN(\ln\frac{K}{N})$.

The generalized Gompertz curve is defined as a solution to the equation

$$\frac{dN}{dt} = rN(\ln\frac{K}{N})^\gamma. \qquad (4.2)$$

Equation (4.2) is a limit case of the generalized logistic equation

$\frac{dN}{dt} = \frac{r}{\beta^\gamma}N[1 - (\frac{N}{K})^\beta]^\gamma$ as $\beta \to 0$.

An inhomogeneous version of Gompertz model (4.2) with distributed Malthusian parameter has a form

$$\frac{dl(t,a)}{dt} = al(t,a)(\ln\frac{K}{N})^\gamma. \qquad (4.3)$$

Again, $l(t, a)$ is a density of the clone, which has the Malthusian parameter equal to $a$. By the same method that was used in ss.2,3 we can show that inhomogeneous Gompertz model also shows the "survival of everybody". Indeed, let us define the "internal population time" $q(t)$ by the equation

$$\frac{dq}{dt} = (\ln \frac{K}{N})^\gamma, \ q(0) = 0.$$

This equation can be written in a closed form (see (A1.8)):

$$\frac{dq}{dt} = (\ln \frac{K}{N(0)M(q)})^\gamma, \ q(0) = 0 \tag{4.4}$$

where $M(q)$ is the mgf of the initial distribution $P(0, a)$ of the parameter $a$.

Equation (4.4) has a unique stable equilibrium $q^*$, which is a solution to the equation $M(q) = K/N(0)$. Then the limit stable population size and the limit distribution of the parameter $a$ are given by the formulas

$$N^* = N(0)M(q^*),$$

$$P^*(a) = P(0; a)\exp(q^* a)/M(q^*). \tag{4.5}$$

Hence, the limit distribution of inhomogeneous Gompertz model (4.3) coincides with the current distribution of the inhomogeneous Malthus model (1.1) at the instant $q^*$; every clone which was presented in the initial time moment will be present in the population forever. So, inhomogeneous Gompertz model also shows non-Darwinian "survival of everybody" and the Exclusion principle does not hold.

## 5. Logistic equation with distributed carrying capacity

Consider the logistic model of inhomogeneous population assuming that the positive parameter $b$, the birth rate, is fixed for the whole population but each clone has its own value of carrying capacity $C$. Dynamics of such a population is described by the equation

$$\frac{dl(t,d)}{dt} = l(t,d)(b - dN(t)) \tag{5.1}$$

where $N$ is the total population size and $d = \frac{1}{C}$ is distributed parameter defining the "death rate".

Let the parameter $d$ takes the values in domain $D$. Let us define the auxiliary variable $q(t)$ by the equations

$$\frac{dq}{dt} = N, q(0) = 0. \tag{5.2}$$

Then

$$l(t, d) = l(0, d)\exp(bt - dq(t)); \tag{5.3}$$

the total size of the population is defined by the formula

$$N(t) = N(0)\exp(bt)M(-q(t)). \tag{5.4}$$

where $M(\delta)$ is the mgf of the initial distribution of the parameter $d$.

Hence, we have reduced inhomogeneous logistic equation (5.1) to a single equation for the keystone variable

$$\frac{dq}{dt} = N(0)\exp(bt) M(-q(t)), q(0) = 0. \tag{5.5}$$

Having the solution of this equation, we can compute the total population size at any moment by formula (5.4); the current distribution of the parameter $d$ is given by the formula

$$P(t, d) = \frac{\exp(-dq(t))}{M(-q(t))} P(0, d). \tag{5.6}$$

Now we can compute all statistical characteristics of interest at any moment given initial distribution $P(0, d)$; for example,

$$E^t[d] = \frac{d \ln(M(\delta))}{d\delta}\bigg|_{\delta=-q(t)}.$$

We can make some conclusions about asymptotical composition of the population at any initial distribution. According to (5.2), $q(t)$ monotonically increases.

**Lemma.** $\lim q(t) = \infty$ as $t \to \infty$.

Assume that $q(t)$ is bounded, $q(t) < K = const$. Then $M(-q) = \int_0^\infty e^{-qx}P(0,x)dx > \int_0^\infty e^{-Kx}P(0,x)dx = M(-K) > 0$, so $\frac{dq}{dt} = N(0)\exp(bt) M(-q(t)) > N(0)\exp(bt) M(-K)$. It follows from here that $q(t) \to \infty$ as $t \to \infty$, in contrast with the assumption.

Let us assume that the minimal possible value $d_{min}$ of the parameter $d$ belongs to the domain of values of the parameter, $d_{min} \in D$.

**Proposition 3**. Inhomogeneous population (5.1) asymptotically consists of a single clone that has the minimal value of the parameter $d$.

Indeed, $\frac{l(t,d_1)}{l(t,d_2)} = \frac{l(0,d_1)}{l(0,d_2)} \exp(-(d_1 - d_2)q(t)) \to 0$ if $d_1 > d_2$

We may come to this and some other assertions about asymptotical behavior of the model by another way. Applying the Price' equation (Price, 1970; see also Rice, 2006) to (5.1), we have

$$\frac{dE^t[d]}{dt} = Cov^t[d, b - dN] = -N(t)Var^t[d] < 0 .\qquad(5.7)$$

It follows from (5.7), that $E^t[d]$ decreases with time until $Var[d] > 0$, i.e. until the population is inhomogeneous; so, $E^t[d]$ tends to minimal possible value of the parameter $d$. It means that asymptotically the population will consist on a clone with $d = d_{min}$.

Asymptotical behavior of the total population size is quite different in the cases $d_{min} > 0$ or $d_{min} = 0$. Namely, in the first case $\lim N(t) = b/d_{min}$ as $t \to \infty$, while in the second case $\lim N(t) = \infty$.

**Example**. Let the initial distribution of the parameter $d$ be exponential in $[0, \infty)$ with mgf $M(q) = 1/(1-q)$. Then the solution to equation (5.5) $q(t) = -1 + \sqrt{(-2 + b + 2e^{bt})/b}$ as $N(0) = 1$, and $N(t) = \frac{be^{bt}}{\sqrt{b(-2+b+2e^{bt})}}$ (see Eq.6.4) increases asymptotically exponentially, see Figure 7.

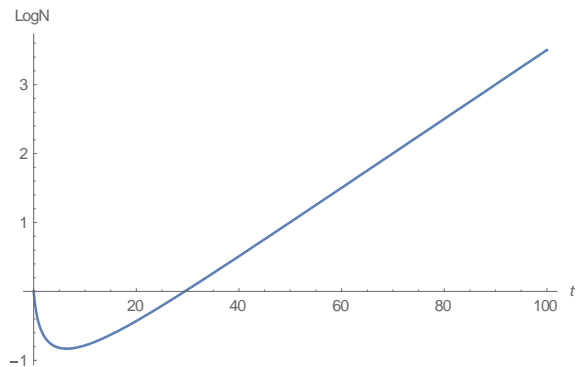

**Figure 7.** Plot of $\log N(t)$, $b = 0.1, d_{min} = 0$; $N(t) = \frac{0.32\, e^{0.1t}}{\sqrt{-1.9+2e^{0.1t}}}$.

In contrast, if the initial distribution of the parameter $d$ is exponential in $[0.01, \infty)$ with mgf $M(q) = e^{0.01q}/(1-q)$, then $\lim N(t) = \frac{b}{d_{min}} = 10$ for $b = 0.1, d_{min} = 0.01$, see Figure 8.

Overall, the logistic model with distributed carrying capacity and fixed Malthusian parameter shows Darwinian "survival of the fittest"; the Exclusion principle is valid for this model.

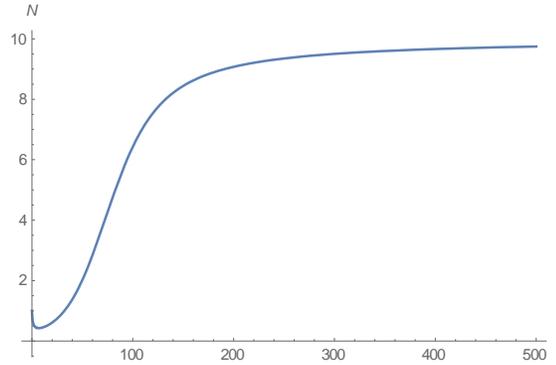

**Figure 8.** Plot of $N(t) = \frac{e^{0.1t - 0.01q(t)}}{1 + q(t)}$; $q(t)$ solves equation (5.2).

6. **Logistic equation with distributed both Malthusian parameter and carrying capacity**

We have shown in the previous sections that the outcomes of evolution of populations described by the inhomogeneous logistic equations can be different. Namely, the logistic-type equations with distributed Malthusian parameter show the survival of everybody, while the equations with distributed carrying capacity demonstrate the survival of the fittest (having the largest carrying capacity).

Let us consider now logistic equations where both Malthusian parameter and carrying capacity are distributed; we can write it in the form of birth-and-death equation:

$$\frac{dl(t;b,d)}{dt} = l(t;b,d)(b - dN). \tag{6.1}$$

This equation was studied in (Ackleh et al., 1999). The authors proved that only "the fittest", i.e. individuals which have the largest value of $b$ and the smallest value of $d$ survive in population in course of time. Similar assertion was proved for more general birth-and-death equations in (Ackleh et al., 2005); corresponding proofs are mathematically rather difficult.

In the rest of this section we apply the HKV method to model (6.1) and give a simple proof of "survival of the fittest". In addition to known results, we found an exact limit distribution of the parameters of this model.

6.1. Consider firstly the case when parameters $b$ and $d$ are independent at the initial time moment and take the values in domains $B$ and $D$ correspondingly. It means that $P(0; b, d) = P_1(0, b) P_2(0, d)$ where $P_1(0, b)$, $P_2(0, d)$ are initial distributions of the parameters $b, d$ correspondingly. Let $M_1(\delta)$, $M_2(\delta)$ be the mgf of distributions $P_1(0, b)$, $P_2(0, d)$.

Introduce the auxiliary variable $q(t)$ as the solution to the Cauchy problem

$$\frac{dq}{dt} = N, q(0) = 0. \tag{6.2}$$

Then the solution to (6.1) reads

$$l(t; b, d) = l(0; b, d)\exp(bt - dq(t)), \tag{6.3}$$

$$N(t) = N(0) \iint_{B,D} \exp(bt - dq(t)) P_1(0, b) P_2(0, d) db dd = N(0) M_1(t) M_2(-q(t)). \tag{6.4}$$

Taking $N(0) = 1$ for simplicity we obtain the closed equation for $q(t)$:

$$\frac{dq}{dt} = M_1(t) M_2(-q). \tag{6.5}$$

**Lemma.** Let $T \leq \infty$ is such that $M_1(t) \to \infty$ as $t \to T$. Then $q(t) \to \infty$ as $t \to T$.

Indeed, if $q(t) < C = const$ for all $t$, then $M_2(-q) > M_2(-C)$ and

$$dq = M_1(t) M_2(-q) dt > M_1(t) M_2(-C) dt.$$

Integrating this inequality and accounting that $q(0) = 0$, we have

$q(t) > M_2(-C) \int_0^t M_1(t) dt$. Now the statement of Lemma is evident.

**Proposition 4.** $\lim_{t \to T} \frac{l(t; b, d)}{l(t; b^*, d_*)} = 0$ for any $b^* \geq b, d_* < d$ or $b^* > b, d_* \leq d$.

Indeed, $\frac{l(t; b, d)}{l(t; b^*, d_*)} = \frac{l(0; b, d)}{l(0; b^*, d_*)} \exp((b - b^*)t - (d - d^*)q(t)) \to 0$ as $t \to T$.

Hence, if $d_{min} \in D$ and $b_{max} \in B$, then only the clone with these values of the parameters survive in course of time. It means the "survival of the fittest". In general case, the distributions of the parameters $d$ and $b$ will concentrate in vicinities of the points $d_{min}$ and $b_{max}$ (including the case $b_{max} = \infty$) but if $d_{min} \notin D$ (or $b_{max} \notin B$) then the total "probability mass" leaves the domain $D$ (or $B$). It means that every clone will be overtaken by another clone in course of time.

6.2. We have shown that the stochastic independence of the birth and death rates in inhomogeneous model (6.1) imply the "survival of the fittest". Now let us consider a general case and do not assume that these rates are independent. We restrict our analysis by the realistic case when each parameter takes only finite number of values. More exactly, we consider a population consisting of $n$ clones (6.1) such that $i$-th clone is characterized by the pair of parameters $(b_i, d_i)$ and is governed by the equation

$$\frac{dl(t;b_i,d_i)}{dt} = l(t;b_i,d_i)(b_i - d_i N) \tag{6.6}$$

Then equations (6.2)-(6.4) take the form:

$$l(t;b_i,d_i) = l(0;b_i,d_i)\exp(b_i t - d_i q(t)), \tag{6.7}$$

$$N(t) = N(0) \sum_{i=1}^{n} P(0;b_i,d_i)\exp(b_i t - d_i q(t)),$$

$$\frac{dq(t)}{dt} = N(t), q(0) = 0.$$

So, in order to solve problem (6.6), we need to study the equation

$$\frac{dq(t)}{dt} = \sum_{i=1}^{n} p_i \exp(b_i t - d_i q(t)) \tag{6.8}$$

where $p_i = P(0;b_i,d_i) \geq 0$ are initial frequencies of the clones, $\sum_{i=1}^{n} p_i = 1$; we assume $N(0) = 1$ for simplicity. The asymptotic behavior of $q(t)$ is studied in Mathematical Appendix 4 with the help of the Method of Newton' diagram. The obtained results allow us to study the asymptotic behavior of current frequencies of the clones:

$$P(t;b_j,d_j) = \frac{p_j \exp(b_j t - d_j q(t))}{\sum_{i=1}^{n} p_i \exp(b_i t - d_i q(t))}, \quad j = 1,\ldots,n \tag{6.9}$$

The problem of our interest is to find the limit values of these probabilities,

$P_j = lim_{t \to \infty} P(t; b_j, d_j)$ and to find those $j$ at which $P_j > 0$.

Before formulating the main result, let us introduce some notations and definitions.

Let $\rho = max_i(b_i/d_i)$ and let $I$ be the set of all indexes such that $\frac{b_i}{d_i} = \rho$. Define a function $f(z) = \sum_{i \in I} p_i z^{d_i}$; let $C \neq 0$ be a solution to the equation $f(z) = \rho$. The function $f(z)$ monotonically increases, so the solution of equation $f(z) = \rho$ exists and is unique.

**Theorem A.** $P_i = p_i C^{d_i}/\rho$, $i \in I$, and $P_j = 0$ for all other $j$.

See Math Appendix 4 for the proof.

The statement of the Theorem can be interpreted as follows. Only the "fittest" clones survive in the population, but the population may have not a single, but several fittest clones. The fittest clone is defined by the condition: $b_i/d_i = \rho$, for all other clones $\frac{b_j}{d_j} < \rho$ and the frequencies of these clones tend to 0 in course of time. By other words, the "fittest" clones are those where the ratio of birth to death coefficients reaches the maximal value possible in the population.

**Example.** Let the population consists of $n = 100$ clones, and the initial distribution is uniform, $p_i = \frac{1}{100}$ for all $i$. Let $b_i/d_i = \rho$ for the first 50 clones, $i = 1, \ldots 50$. Then, according to the Theorem A, the limit frequencies $P_j = 0$ for all $j > 50$ and $P_j > 0$ for all $1 \leq j \leq 50$. Let us assume that $d_i = si$, $s = const$, say, $s = 0.1$. Then $f(z) = \sum_{i \in I} p_i z^{d_i} = \frac{1}{100} \sum_{i=1}^{50} z^{0.1i}$, and $C$ in Theorem A is the root of the equation $\frac{1}{100} \sum_{i=1}^{50} C^{0.1i} = \rho$. Consider two cases: a) $\rho = 0.3$, and b) $\rho = 3$. In case a) $C = 0.8026$; the plot of final distribution is shown on the left panel of Figure 9. In case b) $C = 1.778$; the plot of final distribution is shown on the right panel of Figure 9.

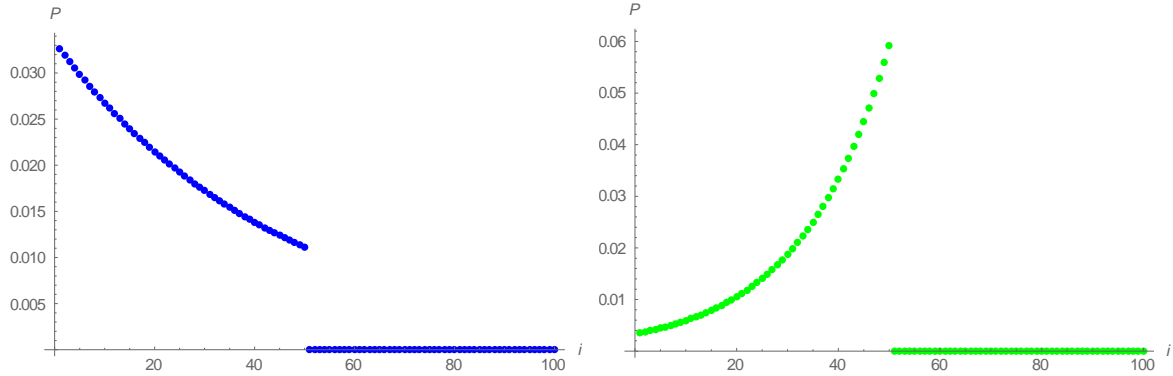

**Figure 9**. Plots of the limit frequences for model (6.6); see explanation in the text of Example.

**7. Dynamics of distributions in inhomogeneous models and the speed of natural selection**

In the previous sections we investigated the dynamics of the total size of various inhomogeneous populations and their final composition in terms of the final distribution of the Malthusian parameter. Here, we will focus on the dynamics of the distribution of the Malthusian parameter, which will allow us to trace the effects and outcomes of natural selection in the aforementioned models.

The current pdf of the Malthusian parameter given its initial distribution for models of the Malthusian type (1.2) is given by the general formula (A1.9). The dynamics of the current pdf is determined by the auxiliary keystone variable $q(t)$ (an "internal time" of the population) as defined by Equation (A1.5). In most cases the function $g(N)$ is positive, so $q(t)$ increases monotonically. A more detailed discussion of the concept of "internal time" within the context of inhomogeneous models can be found in (Karev and Kareva, 2016)).

Dynamics of $q(t)$ depends on the initial distribution of the Malthusian parameter in accordance to Equation (A1.8). The results reported in s.3 highlight an important role of exponential and truncated exponential initial distributions. Notice that it is unrealistic to assume that the Malthusian parameter may have arbitrary large value even with extremely small probability. By this reason, it is more realistic to assume that the initial distribution of the Malthusian parameter is truncated exponential distribution of the form

$$P(0,a) = Ce^{-sa} \text{ for } 0 \leq a \leq B, \tag{7.1}$$

where $C = 1/\int_0^B e^{-sa} sa = \dfrac{s}{1-e^{-Bs}}$ is the normalization constant.

The mgf of the exponential distribution truncated in the interval $[0, B]$ is given by the formula

$$M(t) = \int_0^B e^{at} P(0,a) da = \frac{\left(e^{Bs} - e^{Bt}\right)s}{\left(1-e^{Bs}\right)(t-s)}. \tag{7.2}$$

Then the current pdf is defined by the formula (see (A1.9)):

$$P(t,a) = \frac{e^{a(q(t)-s)}(q(t)-s)}{e^{B(q(t)-s)} - 1}. \tag{7.3}$$

Let us compare the dynamics of truncated exponential distribution with respect to different inhomogeneous models. The simplest models are inhomogeneous versions of the standard Malthusian model, namely density-dependent Malthusian model

$$\frac{dl(t,a)}{dt} = al(t,a) \tag{7.4}$$

and frequency-dependent Malthusian model

$$\frac{dl(t,a)}{dt} = aP(t,a). \tag{7.5}$$

Over time, the distributions in both models tend to become concentrated close to maximal possible value of the distributed parameter $a$, such that $E^t[a] \to B$. Indeed, according to equation (A1.11), $\dfrac{dE^t[a]}{dt} = Var^t[a] > 0$ for D-model (7.4) and $\dfrac{dE^t[a]}{dt} = Var^t[a]/N(t) > 0$ for F-model (7.5). Hence, $E^t[a]$ increases until $Var^t[a] > 0$, i.e. until $E^t[a]$ approach the maximal possible value equal to $B$. Notice that according to equation (A1.3) $\dfrac{dN}{dt} = NE^t[a]$ for D-model and $\dfrac{dN}{dt} = E^t[a]$ for F-model. It follows from here that asymptotically the population size increases exponentially for D-model and linearly for F-model.

Qualitatively, the evolutions of both models are identical up to time change $t \to q(t)$. However, the evolution of F-model (7.5) in real time $t$ is dramatically *slower* compared to the density-dependent model (7.4).

Let us illustrate the difference in the rate of evolution in D- and F-models. A numerical example is given in Figure 11, where the initial distribution is truncated exponential with $s=2$ on the interval $[0,1]$. F-model (7.5) at moment $t$ and D-model (7.4) at moment $t^*$ have identical distributions of the parameter $a$, if $q(t^*)=t$. However, qualitatively, $q(t^*)=10$ if $t^*=1015$, $q(t^*)=15$ if $t^*=86030$, and $q(t^*)=20$ if $t^*=8970000$. As one can see, the rate of evolution is orders of magnitude slower in the frequency-dependent model compared to the density-dependent model.

The reason for such decelerating evolution in the F-model (7.4) compared with D- model (7.5) is that in the F-model, the internal time $q(t) \sim \ln t$. Indeed, in the F-model the population size $N(t)$ is asymptotically linear, $N(t) \sim t$. Hence, according to Equation (A1.5) $\frac{dq}{dt} = \frac{1}{N(t)}$ and

$$q(t) = \int_0^t \frac{du}{N(u)} \sim \ln t.$$

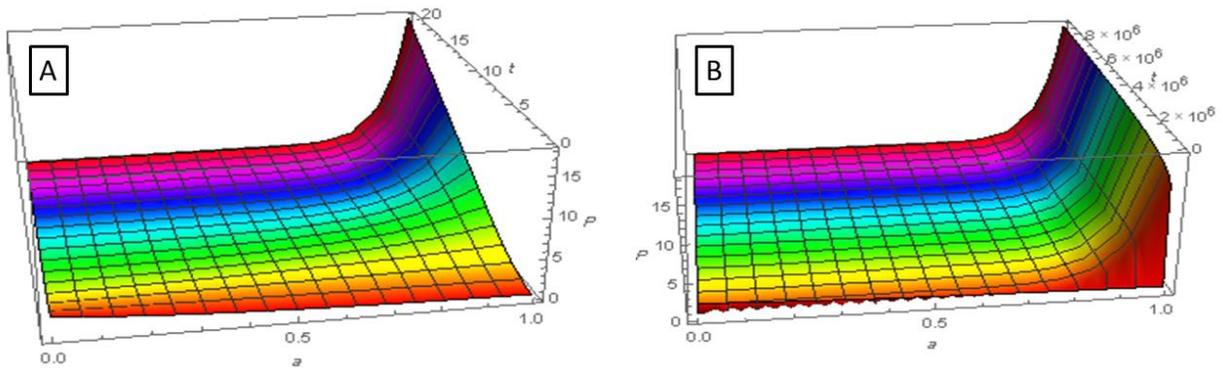

**Figure 10**. Dynamics of distributions of the Malthusian parameter for (A) density-dependent model (7.4) and (B) frequency-dependent model (7.5). Initial distribution is truncated exponential with $s=2$ in the interval $[0,1]$. Distribution of F-model (10) at moment $t=20$ coincides with the distribution of F-model (12) at the moment $t^* = 8970000$.

Overall, both models show Darwinian "survival of the fittest" in the sense that asymptotically, the distribution of the Malthusian parameter $a$ in course of time becomes concentrated in an arbitrarily small vicinity of the maximal possible value of the parameter. The difference is that the rate of evolution in F-model (7.5) decelerates dramatically in comparison to the rate of evolution of the density-dependent model (7.4). Practically, it means that frequency-dependent population is 'much more polymorphic' than the density-dependent population of the same age; frequency-dependent population tends to a monomorphic state but it takes unrealistic time. The Exclusion principle is valid for these models, but only theoretically. The selection of the fittest species requires unrealistic population size for D-model and unrealistic time for F-model.

Inhomogeneous logistic models, as density-dependent (3.2) so frequency-dependent (3.7) demonstrate non-Darwinian "survival of everybody". The reason for this phenomenon is that the "internal time" $q(t)$ for all these models does not increase indefinitely over time, unlike Malthusian-like inhomogeneous models (7.4) and (7.5), but tends to a finite value $q^*$ that solves the equation $\frac{M(q)}{C} = 1$. As a result, the limit distribution $P(t \to \infty, a)$ for the considered logistic-like models given the initial distribution $P(0,a)$ coincides with the distribution of the inhomogeneous density-dependent Malthusian model (7.4) with the same initial distribution $P(0,a)$ taken in the moment $t = q^*$.

Let us emphasize that, similarly to inhomogeneous Malthusian-like models, the evolution of the inhomogeneous frequency-dependent logistic model is much slower than the evolution of the density-dependent logistic model. In Figure 11, one can see the evolution of the initial exponential truncated distribution in $[0,1]$ for logistic D- model (3.2) and F- model (3.7) with $r = \beta = \gamma = 1$.

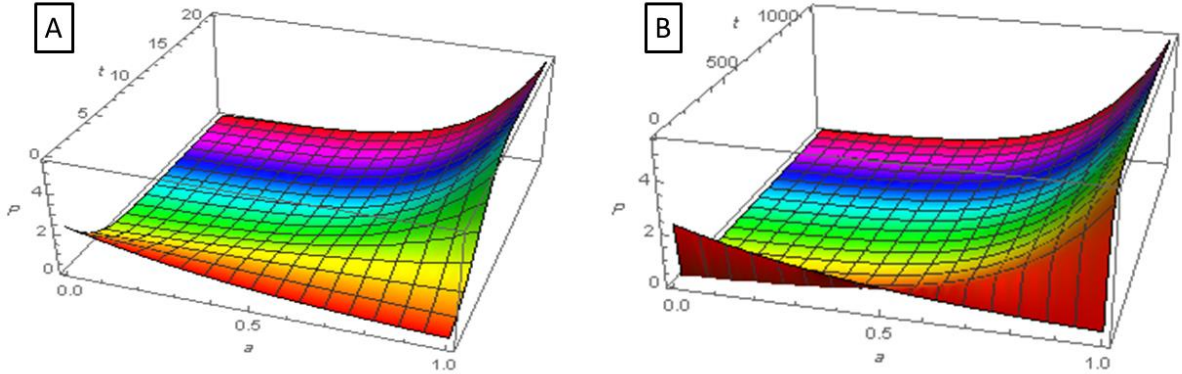

**Figure 11**. Evolution of the distribution of Malthusian parameter in the (A) logistic density-dependent model and (B) logistic frequency-dependent model. The carrying capacity in both models C=100. Initial distribution is truncated exponential in [0,1] with *s*=2.

## Discussion

In this paper we study conceptual mathematical models of natural selection. In order the selection could operate the community or population should be composed from different species or non-identical individuals. So, the models we study are constructed within the frameworks of inhomogeneous density and frequency dependent models. In particular, we consider inhomogeneous versions of classical Malthusian, Gompertzian and different logistic-like models. We show that the outcomes of natural selection in populations described by these models may be different. Specifically, inhomogeneous Malthusian and birth-and-death models show Darwinian "selection of the fittest" while inhomogeneous logistic-like and Gompertzian models show non-Darwinian "survival of everybody". Our approach is based on recently developed HKV (hidden keystone variable) method. According to this method, we introduce an auxiliary variable that can be interpreted as "internal time" of the population. All statistical characteristics of considered inhomogeneous models can be explicitly expressed with the help of the internal time. The outcomes of the natural selection in inhomogeneous populations are determined by the asymptotical behavior of the internal time $q(t)$. Specifically,

1) If $q(t) \to \infty$ as $t \to \infty$ then we have Darwinian *survival of the fittest*;

2) If $q(t) \to \infty$ $as \to t^* < \infty$, then we have non-Darwinian *survival of the common*;

3) If $q(t) \to q^* < \infty$ as $t \to \infty$ then we have non-Darwinian *survival of everybody*.

There are no other options for the population models, which possess the internal time.

The Competitive exclusion principle is often considered a direct consequence of the Darwinian "survival of the fittest". A huge literature is devoted to discussion of this principle (known also as the Gause' principle), that is one of the most important statement in ecology. One of the problem for discussion is: why there are so many different deviations from this principle? We did not discuss in this paper experimental and theoretical aspects of this problem; instead, we concentrated on studying conceptual mathematical models appropriate for modeling of outcomes of natural selection. We have shown that many mathematical models *do not confirm the principle* in its initial strong form as it was formulated by Gause and Hardin; in contrast, the models show coexistence of many species and even "survival of everybody". So, instead of the statement "Complete competitors cannot coexist" we forced to accept the statement" Complete competitors may coexist", at least, in mathematical models. More of that, even the models that demonstrate the Darwinian selection of the fittest, may explain the visible violation of the Exclusion principle: according to frequency-dependent Malthusian model, the evolution is extremely slow and it takes enormous time in order to select the fittest species.

**Mathematical Appendix 1. HKV method and inhomogeneous population models of Malthusian type**

Consider a model of inhomogeneous population

$$\frac{dl(t,a)}{dt} = al(t,a)g(N), \quad N(t) = \int_A l(t,a)da \tag{A1.1}$$

where $g(N)$ is some function of total population size.

Denote $P(t,a) = l(t,a)/N(t)$; the probability density function (pdf) $P(t,a)$ describes the distribution of the parameter $a$ along the population at *t* moment. We suppose that the initial pdf of the Malthusian parameter $a$, $P(0,a)$, is given, and its moment generating function (mgf)

$$M_0(\lambda) = \int_A \exp(\lambda a) P(0,a) da \tag{A1.2}$$

is known.

Denote $E^t[f] = \int_0^\infty f(a) P(t,a) da$ the expected value of $f$ at time *t*. It is known (Hofbauer, Sigmund, 1998) that the population size $N(t)$ satisfies the equation

$$\frac{dN}{dt} = NE^t[a]g(N) \quad (A1.3)$$

and the pdf $P(t,a)$ solves the *replicator equation* of the form

$$\frac{dP(t,a)}{dt} = P(t,a)g(N)(a - E^t[a]). \quad (A1.4)$$

In order to solve the problem (A1.1), let us define formally the "keystone" auxiliary variable $q(t)$ as the solution to the Cauchy problem

$$\frac{dq}{dt} = g(N), q(0) = 0. \quad (A1.5)$$

This equation cannot be solved at this moment, because the population size $N(t)$ is unknown. However, the clone densities and population size can be expressed with the help of the keystone variable $q(t)$:

$$l(t,a) = l(0,a)\exp(aq(t)) = N(0)P(0,a)\exp(aq(t)), \quad (A1.6)$$

$$N(t) = N(0)\int_A \exp(aq(t))P(0,a)da = N(0)M_0(q(t)). \quad (A1.7)$$

Now the equation for the auxiliary variable $q(t)$ can be written in a closed form

$$\frac{dq}{dt} = g(N(0)M_0(q(t))), q(0) = 0. \quad (A1.8)$$

Now we can completely solve the initial problem (A1.1) and corresponding RE (A1.4). The clone densities are given by formulas (A1.6). The population size is given by formula (A1.7) and solves the equation (A1.3).

The current parameter distribution $P(t,a)$ is determined by the formula

$$P(t,a) = \frac{l(t,a)}{N(t)} = P(0,a)\exp(q(t)a)/M_0(q(t)). \quad (A1.9)$$

The mgf of the current distribution $P(t,a)$

$$M_t(\lambda) = E_t[\exp(\lambda a)] = M_0(\lambda + q(t))/M_0(q(t)) \quad (A1.10)$$

The current mean value of the Malthusian parameter solves the equation

$$\frac{dE^t[a]}{dt} = Var^t[a]g(N(t)) \quad (A1.11)$$

and can be computed by the formula

$$E_t[a] = \frac{d\ln M_t}{dq}(q(t)). \quad (A1.12)$$

We refer to the model (A1.1) as a model of Malthusian type, and the variable $q(t)$ can be considered as the "internal time" of the model in the sense that with respect to the new time $q(t)$, each clone grows as if it were independent from other clones. Indeed, making the change of variables $dt \to g(N)dt$, we obtain from (A1.1), (A1.5):

$$\frac{dl(q(t),a)}{dt} = \frac{dl}{dq}\frac{dq}{dt} = al(q(t))g(N).$$

Therefore, $\frac{dl(q,a)}{dq} = al(q,a)$, which is the standard Malthusian model with respect to the "time" $q$, that describes a system of free-growing populations.

**Mathematical Appendix 2. Logistic growth of inhomogeneous population**

We want to construct an inhomogeneous population model whose total size solves the generalized logistic equation

$$\frac{dN}{dt} = rN^\alpha [1 - \left(\frac{N}{C}\right)^\beta]^\gamma . \qquad (A2.1)$$

Let us consider inhomogeneous frequency-dependent model

$$\frac{dl(t,a)}{dt} = raP(t,a)[1 - \left(\frac{N}{C}\right)^\beta]^\gamma \qquad (A2.2)$$

where $a$ is the distributed Malthusian parameter.

Then

$\frac{dN(t)}{dt} = rE^t[a][1 - \left(\frac{N}{C}\right)^\beta]^\gamma$ and we want to have

$$E^t[a] = N(t)^\alpha. \qquad (A2.3)$$

To this end, introduce an auxiliary variable $q(t)$ as a solution to the Cauchy problem

$$\frac{dq}{dt} = [1 - \left(\frac{N}{C}\right)^\beta]^\gamma /N, \ q(0) = 0. \qquad (A2.4)$$

Then $\frac{dl(t,a)}{dt} = ral(t,a)dq/dt$,

$l(t,a) = l(0,a)\exp(raq(t))$,

$$N(t) = N(0)M(rq(t)). \tag{A2.5}$$

Substituting (A2.5) to (A2.4), we obtain a closed equation for $q(t)$. Next,

$$P(t, a) = P(0, a) \exp(raq(t)) / (N(0)M(rq(t))),$$

$$E^t[a] = M'(rq(t))/M(rq(t)).$$

Hence, accounting (A2.3) we arrive to the equation for unknown mgf:

$$\frac{dM(\delta)}{d\delta} = N(0)^\alpha M(\delta)^{\alpha+1}, M(0) = 1.$$

Its solution is

$$M(q) = (1 - \alpha N(0)^\alpha q)^{-1/\alpha}. \tag{A2.6}$$

It is well known that the function $M_0(\delta) = (1 - \beta\delta)^{-\rho}$ at $\beta > 0$ is the mgf of the Gamma-distribution $P(a) = \frac{a^{\rho-1}\exp(-\frac{a}{\beta})}{\beta^\rho \Gamma(\rho)}, > 0$.

It means that the initial distribution of the parameter $a$ in model (A2.2) must be the Gamma-distribution with parameters $\beta = \alpha N(0)^\alpha$ and $\rho = 1/\alpha$, i.e.

$$P(0, a) = \frac{a^{1/\alpha-1}\exp(-\frac{a}{\alpha N(0)^\alpha})}{N(0)\alpha^{1/\alpha}\Gamma(1/\alpha)}. \tag{A2.7}$$

We have proven

**Proposition**. The total population size of the inhomogeneous population (A2.2) with initial Gamma-distribution (A2.7) of the Malthusian parameter $a$ solve the generalized logistic equation (A2.1).

**Mathematical Appendix 3. Multi-factor logistic-like models**

In what follows we assume that the reproduction rates depend on a finite set of regulators of the form

$$G_i(t) = \int_A g_i(a) l(t, a) da, \quad i = 1, \ldots m, \tag{A3.1}$$

where $g_i$ are appropriate functions. In particular, the total population size $N(t)$ corresponds to the function $g(a) \equiv 1$. In the considered above logistic models, $F(t) = 1 - N(t)/C$ for equation (1.3) and $F(t) = N^{\alpha-1}[1 - \left(\frac{N}{K}\right)^\beta]^\gamma$ for equation (3.2), and the population size $N$ was the only regulator.

The current distribution of the population $P(t, a) = l(t, a)/N(t)$, so the mean value of any variable $g_i(a)$ can be expressed as $E^t[g_i] = G_i(t)/N(t)$.

Overall, we assume that the population dynamics can be described by the following "multi-factor" model:

$$\frac{dl(t;a)}{dt} = al(t; a) F(N(t), G_1(t) \ldots G_m(t)). \tag{A3.2}$$

The HKV method for solving of this and more general replicator equations was developed in (Karev, 2010b). According to this method, let us define formally an auxiliary variable $q(t)$ by the equation

$$\frac{dq}{dt} = F(N, G_1 \ldots G_m), \quad q(0) = 0. \tag{A3.3}$$

Then $l(t, a) = l(0, a) \exp(aq(t))$.

Introduce the functional

$$\Phi(r; q) = \int_A r(a) \exp(aq) P(0, a) \, da \tag{A3.4}$$

where $r(a)$ is an arbitrary function (such that the integral in the right hand of (A3.4) exists). Then

$N(t) = N(0) \Phi(1; q(t))$,

$G_i(t) = \Phi(g_i; q(t))$.

Notice that $\Phi(1; q) = M(q)$, where $M$ is the mgf of the initial distribution $P(0, a)$.

Now we can rewrite the equation (A3.3) for auxiliary variable in a closed form:

$$\frac{dq}{dt} = \mathbf{F}(q) \equiv F\big(\Phi(1; q), \Phi(g_1; q) \ldots \Phi(g_m; q)\big), q(0) = 0. \tag{A3.5}$$

So, many-dimensional functional equation (A3.2) is reduced to a single equation (A3.5) for the keystone variable $q$.

In what follows we assume that the equation $\mathbf{F}(q) = 0$ has positive solutions. Let $q^*$ be the minimal solution to $\mathbf{F}(q) = 0$; we suppose that $\frac{d\mathbf{F}}{dq}(q^*) \leq 0$. Then the equilibrium $q^*$ of equation (A3.5) is semi-stable (stable in case of strong inequality).

**Example.** Let $F(x_0, x_1, \ldots x_m) = c(b_0 - x_0)^{k_0}(b_1 - x_1)^{k_1} \ldots (b_m - x_m)^{k_m}$, where $b_i \geq 0, k_i \geq 0$ for all $i$, $c = const > 0$. Then the function $F\big(\Phi(1; q), \Phi(g_1; q) \ldots \Phi(g_m; q)\big) = c(b_0 - \Phi(1; q))^{k_0}(b_1 - \Phi(g_1; q))^{k_1} \ldots (b_m - \Phi(g_m; q))^{k_m}$ satisfies all the conditions.

The limit state of inhomogeneous model (A3.2) as $t \to \infty$ coincides with the current state of the inhomogeneous Malthus model $\frac{dl(q;a)}{dq} = al(q; a)$ at the instant $q^*$. The limit population size, the limit values of regulators and the limit distribution of the parameter $a$ are given by the formulas
$N^* = \Phi(1; q^*)$,

$G_i^* = \Phi(g_i; q^*) = \int_A g_i(a) \exp(aq^*) P(0, a) \, da$,

$P^*(a) = P_0(a)\exp(q^* a)/M_0(q^*)$.

Again, the system remains inhomogeneous at any instant and has a non-trivial limit distribution of the parameter as $t \to \infty$. Every clone that was present initially will be present in the limit stable state. Therefore, inhomogeneous multi-factors model (A3.2) also shows the phenomenon of "survival of everybody" in the population, in contrast to Darwinian "survival of the fittest".

**Mathematical Appendix 4. The Newton diagram method and asymptotic behavior of $q(t)$ for inhomogeneous birth-and-death equation.**

**1.** *Basic equation in the new form*

The function $q(t)$ is defined by equation (6.8):

$$\frac{dq(t)}{dt} = \sum_{i=1}^{n} p_i \exp(b_i t - d_i q(t)), \quad p_i > 0, \sum p_i = 1, \ b_i, d_i \in R_+. \tag{A4.1}$$

The right hand is always positive, hence $q(t)$ monotonically increases; evidently, it cannot tend to a constant, because in this case $dq/dt$ should tends to zero but the right hand of (6.8) increases indefinitely if $q(t)$ tends to a constant. Hence, $lim_{t\to\infty} q(t) = \infty$.

Denote

$$v = Exp(-t) \leftrightarrow t = -\log(v), \quad u = Exp(-q) \leftrightarrow q = -\log(u), \tag{A4.2}$$

$u, v \to 0$ as $t \to \infty$ because $q(t) \to \infty$.

In coordinates $(u,v)$ eq. (A4.1) read

$$\frac{du}{dv} = \frac{du}{dq}\frac{dq}{dt}\frac{dt}{dv} = \frac{u\sum_i p_i v^{-b_i} u^{d_i}}{v} = \frac{u\sum_i p_i u^{d_i} v^{\sigma - b_i}}{v^{\sigma+1}} = \frac{u\sum_i p_i u^{d_i} v^{\sigma_i}}{v^{\sigma+1}} \tag{A4.3}$$

Here $\sigma = \max_{i=1,\ldots,n}\{b_i\}$, $\sigma_i = \sigma - b_i \geq 0$. Without losing of generality we assume that $b_1 \leq b_2 \leq \ldots \leq b_n$, then $0 = \sigma_n \leq \sigma_{n-1} \leq \cdots \leq \sigma_1 = \sigma - b_1 \leq \sigma$.

The condition $q(0) = 0$ implies that $u(0) = 1$.

Equation (A4.3) can be written now in the form of a system

$$\frac{dv}{d\tau} \equiv v' = v^{\sigma+1} = vP(v,u), \tag{A4.4}$$

$$\frac{du}{d\tau} \equiv u' = u\sum_i p_i v^{\sigma_i} v^{d_i} = uQ(v,u)$$

where $(')$ means differentiating by some "dummy" positive variable $\tau$. We have reduced the problem of asymptotical behavior of $q(t)$ to the problem of analysis of the behavior of trajectories of system (A4.4) in a positive neighborhood of the equilibrium point $(v,u) = (0,0)$.

The problem of asymptotical behavior of probabilities $P(t;b_j,d_j)$ is reduced to computation of the values

$$P_j = \lim_{t\to\infty} P(t;b_j,d_j) = \lim_{t\to\infty} \frac{p_j \exp(b_j t - d_j q(t))}{\sum_{i=1}^n p_i \exp(b_i t - d_i q(t))} = \lim_{v,u\to 0} \frac{p_j v^{\sigma_j} u^{d_j}}{\sum_{i=1}^n p_i v^{\sigma_i} u^{d_i}}, \quad j=1,\ldots,n. \quad (A4.5)$$

In order to solve this problem, we apply the method of Newton diagram, developed in (Berezovskaya 1979, 2014; Bruno 2003) and described briefly in the next section.

### 2. Newton diagram method

Let us consider the Kolmogorov`type power vector field $Z(x,y) = \{R(x,y), S(x,y)\}$ given by system of differential equation:

$$x' = x \sum_{\mu,\nu} r_{\mu\nu} x^\mu y^\nu = xR(x,y), \quad \mu,\nu \geq 0 \quad (A4.6)$$

$$y' = y \sum_{\mu,\nu} s_{\mu\nu} x^\mu y^\nu = yS(x,y),$$

*Definition* 1. A set $M = \{(\mu,\nu): |r_{\mu\nu}| + |s_{\mu\nu}| \neq 0\}$ is the *support* of vector field (A4.6); $(r_{\mu\nu}, s_{\mu\nu})$ is the *vector-coefficient* of the point $(\mu,\nu) \in M$.

*Definition* 2. *Newton diagram* (ND) $\Gamma$ is the convex hull of $\{(\mu,\nu) + R_+^2\}$ if $(\mu,\nu) \in M$; $\Gamma$ may consist of one vertex $\gamma^{(0)}$ or is a polygonal line, which consists of edges $\gamma \in \Gamma$ together with their vertexes, see Fig.A4.1.

*Definition* 3 (a). *Index* of the *vertex* $(\mu,\nu)$ is the value $\beta \equiv \beta(\mu,\nu) = \frac{s_{\mu\nu}}{r_{\mu\nu}}$ if $r_{\mu\nu} \neq 0$ and $\beta = \infty$ if $r_{\mu\nu} = 0$.

(b) *Index* of the *edge* $\gamma$ is the value $\alpha \equiv \alpha(\gamma) = \frac{\mu_1 - \mu_2}{\nu_2 - \nu_1} > 0$ if the points $A_1(\mu_1,\nu_1)$, $A_2(\mu_2,\nu_2)$, $\mu_1 \neq \mu_2, \nu_1 \neq \nu_2$ belong to the edge $\gamma$.

**Remark** A4.1. The edge index $\alpha(\gamma)$ is equal to the slope of the line $l$ passing throw the points $A_1, A_2$ with negative direction of the ordinate axis.

Let $M$ be the support containing $n \geq 1$ points and $\Gamma$ be the Newton diagram of vector field (A4.6). Denote $Z_\gamma(R_\gamma, S_\gamma\}$ the truncation of (A4.6) to the edge $\gamma$; $Z_\gamma$ is the vector field of the form (A4.6), where summation is performed over $(\mu,\nu) \in M_\gamma \equiv M \cap \gamma$.

*Definition* 4. Vector field $Z(R, S)$ is *non-degenerate* if

1) index $\beta$ of any vertex of $\Gamma$ does not equal to indexes of edges adjacent to this vertex;

2) for any edge $\gamma$ the functions $R_\gamma(1, z), S_\gamma(1, z)$ have no common non-zero roots;

3) the function $F_\gamma(z) = -\alpha R_\gamma(1, z) + S_\gamma(1, z)$, $\alpha = \alpha(\gamma)$ has no multiple non-zero roots.

Evidently, that non-degenerate Kolmogorov`type vector field $Z$ has trivial orbits $x = 0$ and $y = 0$, hence the isolated singular point $O$ is not monodromic (i.e., is not a center or a focus).

The following Theorem describes asymptotical behavior of orbits in a small neighborhood of singular point $O$ (Berezovskaya 1979, 2014). Let us call to *O-orbit* any orbit $(x(t), y(t))$ such that $(x(t), y(t)) \rightarrow (0,0)$ as $t \rightarrow \pm\infty$.

**Theorem A4.1.** *Let Z be non-degenerate Kolmogorov`type vector field. Then all non-trivial O-orbits of Z have power asymptotics*

$$y = K^* x^\rho (1 + o(1)), K^* \neq 0, \rho > 0, x \rightarrow 0 \qquad (A4.7)$$

*where the power $\rho$ is equal to the index $\beta$ of a vertex $\in \Gamma$ or to the index $\alpha$ of an edge $\in \Gamma$;*

i) $\rho = \beta$ *if and only if $\tilde{\alpha} < \beta < \hat{\alpha}$ where $\tilde{\alpha}, \hat{\alpha}$ are indexes of edges adjacent to the vertex ($\mu$, $\nu$) having index $\beta$; $\tilde{\alpha} = 0$ if vertex ($\mu, \nu$) belongs to ordinate axis, i.e., $\mu = 0$, and $\hat{\alpha} = \infty$ if vertex ($\mu, \nu$) belongs to abscise axis, i.e., $\nu = 0$. The coefficient $K^*$ in formula (A4.7) is an arbitrary constant;*

ii) $\rho = \alpha = \alpha(\gamma)$ *if and only if the function $F_\gamma(z)$ has a root $z = K^*$.*

*Remark* A4.2. Several asymptotics defined by the Theorem may exist simultaneously.

### 3. Asymptotics of orbits of system (A4.4)

Let us apply the ND method to analysis of the asymptotic behavior of system (A4.4)

$$v' = v^{\sigma+1} = vP(v, u), \qquad u' = u \sum_i p_i v^{\sigma_i} u^{d_i} = uQ(v, u), \qquad (A4.8)$$

$P(v, u) = v^\sigma$, $Q(v, u) = \sum_i p_i v^{\sigma_i} u^{d_i}$ and $0 = \sigma_n \leq \sigma_{n-1} \leq \cdots \leq \sigma_1 \leq \sigma \equiv \sigma_0$.

The support $M$ of system (A4.4), (A4.8) consists of $n+1$ points: $A_0 = A_0(\sigma_0, 0)$ with vector-coefficient (1,0), and $A_i = A_i(\sigma_i, d_i)$ with vector-coefficients $(0, p_i), i = 1,..n$. Let us construct the Newton diagram $\Gamma$ of the system (see Figure A4.1).

Between all the points that are placed on the ordinate axes $d$ let us choose the point $D(0, \min(d_j)) = D(0, d)$. Evidently, the points $A_0, D$ are vertexes of $\Gamma$. The line $l$ that connects these two points has the slope $\alpha(l) = \frac{\sigma_0}{d}$ with the negative direction of ordinate axis (see Remark A4.1). The lines $l^i$ connecting the point $A_0(\sigma_0, 0)$ and the points $A_i(\sigma_i, d_i), i = 1,..,n$, have slopes $\alpha^i = \frac{\sigma_0 - \sigma_i}{d_i}$.

If $\alpha(l) = \frac{\sigma_0}{d} \leq \alpha^i, i = 1,..,n$ then ND has *unique* edge $\gamma_1$, and $\alpha_1 \equiv \alpha(\gamma_1) = \frac{\sigma_0}{d}$ is its index. Points $A_i(\sigma_i, d_i)$ such that $\alpha^i \equiv \frac{\sigma_0 - \sigma_i}{d_i} = \frac{\sigma_0}{d} = \alpha_1$ also belong to the edge $\gamma_1$ (and are enumerated $A_{1i}(\sigma_{1i}, d_{1i}), i \epsilon I_{\gamma_1}$ whereas other points of support $M$ are placed upper this edge and so don't belong to $\Gamma$ (see Figure A4.1- a).

If the minimal slope $\alpha_1 = \min_i \alpha^i$ is achieved by the line connecting the points $A_0$ and $C \equiv C(\sigma_c, d_c)$ where $\sigma_c \neq 0$ then the Newton diagram $\Gamma$ has the edge $\gamma_1$ whose vertexes are the points $A_0(\sigma, 0)$ and $C(\sigma_c, d_c)$ and $\alpha_1 \equiv \alpha(\gamma_1) = \frac{\sigma - \sigma_c}{d_c}$. In this case $\Gamma$ has at least two edges.

The case when ND of system (A4.4) has exactly 2 edges, $\gamma_1, \gamma_2$, is presented in Figure A4.1- b. Here, the edge $\gamma_1$ bounded by vertices $A_0, C$ has index $\alpha_1 = \frac{\sigma_0 - \sigma_c}{d_c}$, and the edge $\gamma_2$ bounded by vertices $C$ and $D$ has index $\alpha_2 = \frac{\sigma_c}{d - d_c} < \alpha_1$. Generally, the Newton diagram $\Gamma$ can have some more edges. Thus we prove the following statement.

***Proposition A4.1.*** *The Newton diagram $\Gamma$ of system (A4.4) always contains the edge $\gamma_1$ with the vertex $A_0(\sigma_0, 0)$; it can also contain the edges $\gamma_2, ... \gamma_r, r \geq 2$. All edges compose a polygonal line connecting the vertexes $A_0(\sigma_0, 0)$ and $D(0, d)$.*

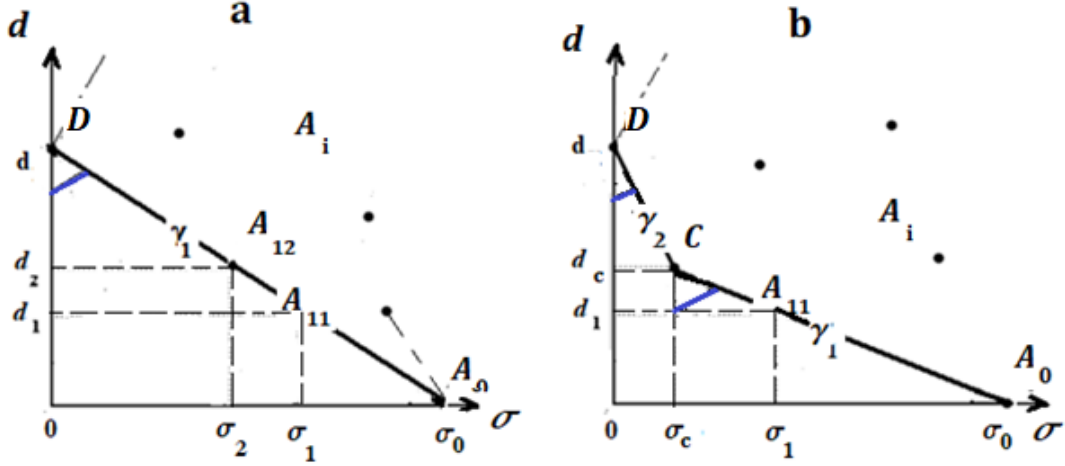

**Figure** A4.1. The Newton diagram is the convex polygonal line passing through the points $A_0(\sigma_0, 0), D(0, d)$ such that each $A_i$ lies above or on it. (**a**) The diagram consists of one edge $\gamma_1$ whose index is $\alpha_1$; (**b**) the diagram consists of two edges $\gamma_1, \gamma_2$ whose indexes are $\alpha_1 > \alpha_2$.

Now we can apply Theorem A4.1 to system (A4.8) to find asymptotical behavior of orbits of (A4.8).

***Proposition A4.2.*** *Orbits of system (A4.8) have only one non-trivial asymptotics*

$$u = Kv^\rho(1 + o(1)), v \to 0. \tag{A4.9}$$

where $K > 0$, $\rho = \alpha(\gamma)$, and $\gamma$ is the edge of Newton diagram of (A4.4) that has the vertex $(\sigma_0, 0)$.

*Proof.* Due to Theorem A4.1 a power $\rho$ of asymptotics (A4.4) can be equal only to a vertex or an edge index. The vertex $(\sigma_0, 0)$ of Newton diagram has index $\beta = \beta_0 = 0$. Index of any point $A_i(\sigma_i, d_i) \in M$, $d_i > 0$, $i = 1, \ldots n$ (including point $D(0, d)$) is equal to $\infty$ (according to definition 3(a)). Thus, system (A4.8) has no orbits with power asymptotics (A4.9) that correspond to vertexes of $\Gamma$.

Next, let us study the asymptotics (A4.9) that correspond to edges of diagram $\Gamma$. Let us suppose that the Newton diagram of system (A4.4) has only one edge $\gamma = \gamma_1$ with the vertexes $A_0(\sigma, 0)$, $B(0, d)$; $\alpha = \frac{\sigma_0}{d}$ is the edge index. Then $P_\gamma(v, u) = v^{\sigma_0}$, $Q_\gamma(v, u) = \sum_{j \in I(\gamma_1)} p_j v^{\sigma_j} u^{d_j}$ and $P_\gamma(1, z) = 1, Q_\gamma(1, z) = \sum_{j \in I(\gamma_1)} p_j u^{d_j}$. Because $p_j \geq 0 \; for \; j \in I(\gamma_1), p_0 \neq 0$ then function $F_\gamma(z) = -\alpha P_\gamma(1, z) + Q_\gamma(1, z) = -\frac{\sigma_0}{d} + \sum_{j \in I(\gamma_1)} p_j u^{d_j}$. This function has only one positive root, say, $z = K^{**}$, because $\sum_{j \in I_\gamma} p_j z^{d_j}$ monotonically increases with $z$. Applying Theorem A4.1 we can state that the system has orbits with asymptotics (A4.9) $u = Kv^\rho(1 + o(1))$, where $\rho = \alpha = \frac{\sigma_0}{d}, K = K^*$, and that there are no orbits with other positive asymptotics with $\rho = \alpha$.

Now consider the case when the Newton diagram contains two edges, $\gamma_1$ and $\gamma_2$. Then the edge $\gamma_1$ has vertexes $A_0(\sigma, 0), C(\sigma_c, d_c)$ and the index $\alpha(\gamma_1) = \alpha_1$, the edge $\gamma_2$ has vertexes $C(\sigma_c, d_c), B(0, d), d > d_c$ and index $\alpha(\gamma_2) = \alpha_2; \alpha_1 > \alpha_2$. As in previous case, the function $F_{\gamma_1}(z) = \alpha_1 - \sum_{j \in I(\gamma_1)} p_j z^{d_j}$ has only one positive root. The truncation of system (A4.4), (A4.8) into edge $\gamma_2$ is given by $P_{\gamma_2}(v, u) \equiv 0, Q_{\gamma_2}(v, u) = \sum_{j \in I(\gamma_2)} p_j v^{\sigma_j} u^{d_j}$, so the function $F_{\gamma_2}(z) = \sum_{j \in I(\gamma_2)} p_j z^{d_j}$ evidently has no positive roots (because all $p_j$ are positive). Due to Theorem A4.1 the orbits of vector field (A4.4) have no positive asymptotics (A4.9) with $\rho = \alpha_2$. Similarly, if the Newton diagram contains $r > 2$ edges, then the functions $F_{\gamma_k}(z) = -\sum_{j \in I(\gamma_k)} p_j z^{d_j}$, $k = 2, \dots r$ have no positive roots, and so positive orbits of (A4.4) have no asymptotics (A4.9) with the powers $\rho = \alpha_k, k > 1$.

*The proposition is proven.*

Notice now that, by definition, $\alpha(\gamma_1) = \min_i \frac{\sigma_0 - \sigma_i}{d_i} = \max_i (\frac{b_i}{d_i})$, hence $\rho = \alpha(\gamma_1) = \max_i (\frac{b_i}{d_i})$.

Then, Propositions A4.1 and A4.2 imply

**Theorem A4 2.** *Orbits of system* (A4.4), (A4.8) *which tend to* $(u = 0, v = 0)$ *from positive initial values have unique non-trivial asymptotics (A4.9)* $u = Kv^\rho(1 + o(1))$. *Here* $\rho = \max_i(\frac{b_i}{d_i})$ *and $K$ is a positive root of the function* $F_\gamma(z) = -\rho + \sum_{j \in I_\gamma} p_j z^{d_j}$.

## 2.4. Asymptotics of probabilities

In order to compute the limiting probabilities $P_j = P(t; b_j, d_j)$ given by formula (A4.5) we need the following estimations.

Let $\Phi(x, y) = \sum_{(\mu,\nu) \in M} f_{\mu,\nu} x^{\mu} y^{\nu}$, $f_{\mu,\nu} \neq 0$, $\Phi(0,0) = 0$, $M = \{(\mu,\nu)\}$ be a support of the function $\Phi$, and $\Gamma$ be the corresponding Newton diagram, i.e., the convex hull of the points $\{(\mu,\nu) + R_+^2\}$ for $(\mu,\nu) \in M$. Let $\Phi_\gamma(x, y) = \sum_{(\mu,\nu) \in M \cap \gamma} f_{\mu,\nu} x^{\mu} y^{\nu}$ be the truncation of $\Phi(x, y)$ to an edge $\gamma$ having the index $\alpha$.

**Lemma A4.1.** *There exist non-negative constants $\lambda$, $\delta$ such that for arbitrary constant $z$*

$$\Phi(x, zx^\alpha) = x^\lambda (\Phi_\gamma(1, z) + x^\delta \varphi(x, z)),$$

*where* $\Phi_\gamma(1, z) = \sum_{(\mu,\nu) \in M \cap \gamma} f_{\mu,\nu} z^\nu$ *and* $\varphi(0,0) = 0$.

*Proof.* Let us write the function $\Phi(x, y)$ in the form

$$\Phi(x, y) = \Phi_\gamma(x, y) + \tilde{\Phi}(x, y) = \sum_{(\mu,\nu) \in M \cap \gamma} f_{\mu,\nu} x^\mu y^\nu + \sum_{(\mu,\nu) \in M \setminus \gamma} f_{\mu,\nu} x^\mu y^\nu.$$

The edge $\gamma$ belongs to the line $\mu + \alpha\nu = c = const$ if $(\mu,\nu) \in M \cap \gamma$, and $\mu + \alpha\nu > c$ if $(\mu,\nu) \notin M \cap \gamma$. Then

$$\Phi_\gamma(x, zx^\alpha) = \sum_{\mu+\alpha\nu=c} f_{\mu,\nu} x^{\mu+\alpha\nu} z^\nu = x^c \sum_{\nu=(c-\mu)/\alpha} f_{\mu,\nu} z^\nu,$$

$$\tilde{\Phi}(x, zx^\alpha) = \sum_{\mu+\alpha\nu>c} f_{\mu,\nu} x^{\mu+\alpha\nu} z^\nu = x^c \sum_{\mu,\nu} f_{\mu,\nu} x^{\delta_{\mu\nu}} z^\nu$$

where $\delta_{\mu\nu} > 0$. Thus, taking $\lambda = c$ and $0 < \delta < \min_{(\mu,\nu) \in (M - M \cap \gamma)}(\delta_{\mu\nu})$ we finish the proof.

Asymptotical values of probabilities are defined by formulas (A4.5):

$$P_j = \lim_{v,u \to 0} \frac{p_j v^{\sigma_j} u^{d_j}}{\sum_{i=1}^n p_i v^{\sigma_i} u^{d_i}}, \; j = 1,\ldots,n \text{ where } \sigma_i = \sigma - b_i \geq 0, \; \sigma = \max_{i=1,\ldots,n}\{b_i\}.$$

Let us present the function $G(u,v) = \sum_{i=1}^n p_i v^{\sigma_i} u^{d_i}$ as the sum of two terms, the first one is the truncation of $G(u,v)$ to the edge $\gamma_1$ having index $\rho$, and the second one contains all other summands:

$$\sum_{i=1}^n p_i v^{\sigma_i} u^{d_i} = \sum_{(\sigma_i,d_i) \in M \cap \gamma_1} p_i v^{\sigma_i} u^{d_i} + \sum_{(\sigma_i,d_i) \in M \setminus \gamma_1}^{n_1} p_i v^{\sigma_i} u^{d_i} = \sum_{i \in I(\gamma_1)} p_i v^{\sigma_i} u^{d_i} + \sum_{i \notin I(\gamma_1)} p_i v^{\sigma_i} u^{d_i}.$$

Recall that, by definition, the index $\alpha(\gamma_1) = \frac{\sigma_0 - \sigma_i}{d_i} = \rho$, hence $\sigma_i + \rho d_i = \sigma$ if $i \in I(\gamma_1)$; for any $j \bar{\in} I(\gamma_1)$ we have $\frac{\sigma_0 - \sigma_j}{d_j} < \rho$, hence, $\sigma_j + \rho d_j = \sigma + \delta_j$, $\delta_j > 0$ if $j \bar{\in} I(\gamma_1)$.

Using power asymptotics (A4.9) and Lemma A4.1 we can write for $j = 1, \ldots n_1$:

$$\frac{p_j v^{\sigma_j} u^{d_j}}{\sum_{i=1}^{n} p_i v^{\sigma_i} u^{d_i}} = \frac{p_j v^{\sigma_j + \rho d_j} K^{*d_j}(1+o(1))}{\sum_{i \in I(\gamma_1)} p_i v^{\sigma_i + \rho d_i} K^{*d_i}(1+o(1)) + v^{\delta} \sum_{i \notin I(\gamma_1)} p_j v^{\sigma_i + \rho d_i - \delta} K^{*d_i}(1+o(1))} =$$

$$\frac{v^{\sigma} p_j K^{*d_j}(1+o(1))}{v^{\sigma}(\sum_{i \in I(\gamma_1)} p_i K^{*d_i} + v^{\delta} \sum_{i \notin I(\gamma_1)} p_j K^{*d_i} v^{\delta_j})(1+o(1))} \to \frac{p_j K^{*d_j}}{\sum_{i \in I(\gamma_1)} p_i K^{*d_i}}.$$

So,

$$P_j = \lim_{v,u \to 0} \frac{p_j v^{\sigma_j} u^{d_j}}{\sum_{i=1}^{n} p_i v^{\sigma_i} u^{d_i}} = \frac{p_j K^{*d_j}}{\sum_{i \in I(\gamma_1)} p_i K^{*d_i}}, j \in I(\gamma_1) \text{ and } \sum_{j \in I(\gamma_1)} P_j = 1.$$

As $\sum_{i=1}^{n} P_i = 1$, then $P_j = 0$ for all $j \bar{\in} I(\gamma_1)$.

We have proven

**Theorem A4.3.** *Let* $\rho = max_i(\frac{b_i}{d_i})$, $S = \{(b_i, d_i): \frac{b_i}{d_i} = \rho\}$, *and* $I = \{i: (b_i, d_i) \in S\}$; *let* $K^*$ *be the single positive solution of the equation* $\sum_{i \in I} p_i z^{d_i} = \rho$. *Then* $P_i = \frac{p_i K^{*d_i}}{\sum_{j \in I} p_j K^{*d_j}} > 0$ *for* $i \in I$ *and* $P_j = 0$ *for all* $j \bar{\in} I$.